# Method for classifying a noisy Raman spectrum based on a wavelet transform and a deep neural network

Liangrui Pan[1], (Student Member, IEEE), Pronthep Pipitsunthonsan [1], (Student Member, IEEE), Chalongrat Daengngam[2], Sittiporn Channumsin[3], Suwat Sreesawet[3], Mitchai Chongcheawchamnan[1], (Senior Member, IEEE)

[1]Faculty of Engineering, Prince of Songka University, Songkhla, 90110 Thailand
[2]Faculty of Science, Prince of Songka University, Songkhla, 90110 Thailand
[3]Geo-Informatics and Space Technology Development Agency (GISTDA), Chonburi 20230, Thailand

Corresponding author: Mitchai Chongcheawchamnan (mitchai.c@psu.ac.th).

This work is funded by Science, Reserach and Innovation Promotion Fund (Grant No: 1383848).

**ABSTRACT** This paper proposes a new framework based on a wavelet transform and deep neural network for identifying noisy Raman spectrum since, in practice, it is relatively difficult to classify the spectrum under baseline noise and additive white Gaussian noise environments. The framework consists of two main engines. Wavelet transform is proposed as the framework front-end for transforming 1-D noise Raman spectrum to two-dimensional data. This two-dimensional data will be fed to the framework back-end which is a classifier. The optimum classifier is chosen by implementing several traditional machine learning (ML) and deep learning (DL) algorithms, and then we investigated their classification accuracy and robustness performances. The four MLs we choose included a Naive Bayes (NB), a Support Vector Machine (SVM), a Random Forest (RF) and a K-Nearest Neighbor (KNN) where a deep convolution neural network (DCNN) was chosen for a DL classifier. Noise-free, Gaussian noise, baseline noise, and mixed-noise Raman spectrums were applied to train and validate the ML and DCNN models. The optimum back-end classifier was obtained by testing the ML and DCNN models with several noisy Raman spectrums ($10 - 30$ dB noise power). Based on the simulation, the accuracy of the DCNN classifier is 9% higher than the NB classifier, 3.5% higher than the RF classifier, 1% higher than the KNN classifier, and 0.5% higher than the SVM classifier. In terms of robustness to the mixed noise scenarios, the framework with DCNN back-end showed superior performance than the other ML back-ends. The DCNN back-end achieved 90% accuracy at 3 dB SNR while NB, SVM, RF, and K-NN back-ends required 27 dB, 22 dB, 27 dB, and 23 dB SNR, respectively. In addition, in the low-noise test data set, the F-measure score of the DCNN back-end exceeded 99.1% while the F-measure scores of the other ML engines were below 98.7%.

## I. INTRODUCTION

Raman spectroscopy is a material characterisation method widely used in industrial process controls, planetary exploration, homeland security, life science, geological field investigation, and laboratory material research [1]. By identifying the Raman spectrum of a small number of substances, an accurate label of the substance can be obtained [2]. For example, in the detection of minerals in the field, we may only sample all the minerals and then perform experimental analysis on them. It is necessary to perform pre-processing for obtaining Raman spectra, such as using

Raman spectroscopy to check the composition of chemical substances and implement statistical classification methods. Preferably, a rapid and accurate classification algorithm is required when dealing with a large Raman spectrum set. Nowadays, there are many chemical/biochemical molecular structure databases for researchers to access, such as the FT-Raman spectra database [3], an e-VISART database [4], a biomolecule database [5], and an explosive compound database [6]. These databases contain a large amount of raw and processed Raman data for Raman spectroscopy application.

For practical Raman spectrum data, noise signals in the spectrum can originate from several sources such as the fluorescence process, material density, external light source, charge-coupled device receivers, external charge amplifiers, and environmental noise. In the signal processing aspect, we need to reduce these noises before performing classifying. Without the process to reduce such noises, the Raman classification accuracy will deteriorate. Hence, several methods were proposed; for example, baseline correction [7] and surface enhancement [8], to name a few. Although some algorithms have been improved based on these methods, they are still challenging for fully automatic processing of Raman spectra. The random noise affects the peaks and the sub-peaks of the Raman spectrums. This causes the extraction of spectrum peaks and subpeaks difficulty, and finally reduces the classification accuracy.

In the past few decades, there have been many automatic baseline correction algorithms based on the original Raman spectrum such as the Least Square method, Asymmetric Least Square method (AsLS), and the Penalty Least square method (PLS). Zhang et al. proposed an adaptive iterative reweighted Penalty Least Squares (airPLS) algorithm without prior information, such as user intervention and peak detection [7]. He et al. proposed a baseline correction method of Raman spectrum correction with the Improved Asymmetric Least Squares (IAsLS) [9]. The algorithm estimates the original spectral line by a polynomial fitting method. Compared with AsLS, the root means-square-error of IAsLs was reduced by 16 times, and the baseline can be automatically subtracted. Beak et al. proposed a weighted airPLS method by using the generalised logic function based on the baseline correction of the PLS, which estimates the noise level iteratively and adjusts the weight accordingly [10].

Apart from baseline correction, noise level also poses severe effects on the peak characteristics of the Raman spectrum. Ehrentreich et al. proposed a wavelet transform (WT) to identify the peak value through the first level detail coefficient [11]. The position of the spike can be projected from detail to approximation and then to the appropriate position of the original spectrum. After the peak is determined, these regions will be replaced by subtraction. Barclays et al. proposed a discrete WT for spectrum smoothing and denoising to remove small-amplitude components independent of position in the transform domain [12]. The method has excellent performance in an extended dynamic range. Guo et al. proposed a method combining the Mexican-Hat wavelet and average algorithm to extract the Raman signal from high and low spectral noise [13]. This method has the characteristics of small relative errors of spectrum intensity and spectrum width. These previous works show that the WT can effectively be applied to Raman spectroscopy.

Based on ML algorithms, several researchers have proposed a better classification and recognition model, from a simple perceptron to a vast artificial neural network. Several researchers aimed to develop ML algorithms for classification problems. These include the decision tree (DT), Naive Bayes (NB), K-Nearest-Neighbor (K-NN), and Support Vector Machine (SVM), etc [1], [14]–[16]. DT (a graphic method of intuition using probability analysis [14]) is a decision analysis method to calculate the probability that the expected value of the net present value is greater than or equal to zero. The DT is formed according to the probability of various situations. The greedy algorithm is applied to build the DT by only considering the condition of the maximum purity difference as the segmentation point [17]. The construction of the DT is a recursive process. On the other hand, SVM aims to find a separate hyperplane in the feature space, which divides different data instances into various labels to achieve classification [16]. This algorithm does not make any assumptions about the distribution of the original data set, so it is widely used in biomedical engineering, chemical materials, and physical spectra. Effendi et al. evaluated the ability of near-infrared Raman spectroscopy combined with SVM to improve the classification of different histopathological groups in tissues [18]. Two types of SVM (i.e. C-SVM and v-SVM) with three kernel functions, the linear, polynomial and Gaussian radial basis function (RBF), are used in combination with Principal Component Analysis to develop an effective algorithm for classifying the Raman spectra of different colonic tissues. N. H. Othman et al. evaluated the ability to combine near-infrared Raman spectroscopy with SVM to improve the multi-class classification of different histopathological groups in tissues. The diagnostic accuracy of 99.9% for multi-class classification was obtained [4]. Its disadvantages are that the efficiency is not very high when there are many observation samples. Other disadvantages include that there are no general solutions for nonlinear problems and that it is sensitive to missing data.

Traditional ML classifiers such as NB, RF, K-NN, and SVM were applied to Raman spectroscopy [18]–[21]. It was shown that NB has fast convergence. Julio et al. used Raman spectroscopy and Bayesian classifiers to classify breast biopsies of healthy and cancerous tissues with 100% accuracy [19]. The main disadvantage of the NB algorithm is its feature redundancy. For the RF algorithm, it is simple to understand and interpretable. The algorithm was applied to analyse complex Surface Enhanced Raman Scattering (SERS) data to obtain accurate and complex interpretations based on previous knowledge about the available SERS signals [20]. One of its disadvantages is that it does not support online learning, so after the arrival of new samples, the decision tree, which is the RF algorithm structure, needs to be rebuilt. Another drawback of the RF algorithm is its overfitting problem. On the other hand, the K-NN algorithm is suitable for automatic classification of class domains with a large sample size. In [21] K-NN classified SERS data it was able to early detect dengue fever with a classification accuracy of 82.14%. The main disadvantage of this algorithm is that it is easy to generate misclassification for those class domains with small sample sizes where the output is also not interpretable. SVM can solve high-dimensional problems such as large-scale feature space, small sample sizes and interaction of nonlinear features. N. H. Othman et al. evaluated the ability to combine near-infrared Raman spectroscopy with SVM to improve the multi-class classification of different histopathological groups in tissues. A diagnostic accuracy of 99.9% for multi-class classification was obtained [18]. Its disadvantages are that its efficiency is not very high when there are many observation samples. Other disadvantages include no general solution for nonlinear problems and that it is sensitive to missing data.

Unlike the ML algorithms, the goal of the deep learning (DL) method is to learn the feature level of high-level features composed of low-level feature groups [22]. Applying the DL method to NN, various deep neural networks (DNNs) have been proposed and achieved more than 90% accuracy, which is better than the traditional NNs. Weng *et al*. applied the migration network framework to biomedical engineering and used the coherent anti-Stokes Raman scattering image to diagnose lung cancer automatically [23]. Using this model to analyse other cancer cells, the accuracy of cancer cell image recognition is 89.2%, confirming that the DNN is a powerful image processing technology. Natalia *et al*. constructed a multi-level DL framework, the core of which is the unsupervised NN and a group of supervised NNs [24]. The accuracy of this particular DNN is 85% compared with a convolution neural network (CNN) in the classification of land cover and crop types in multi-band and multi-source satellite images.

In the signal theory aspect, the Raman spectrum is a one-dimensional (1-D) signal, hence 1-D CNN was proposed to identify a spectrum peak from a noisy Raman spectrum. M. Fukuhara et al. used a digitally generated Lorentz spectrum to determine the optimal filter size (close to line width) and the number of filters to extract Raman peaks. However, this method has many steps and the extracted peaks were partially missing. The recognition accuracy of 1-D CNN is very low in a relatively large noise environment (noise close to the sub-peak) [25]. Consequently, we propose a Raman spectrum classification algorithm based on a two-dimensional deep convolution neural network (2-D DCNN). WT is proposed to transform 1-D noisy Raman spectra to a 2-D scape map [19]. All spectrum information and noise information on the noisy Raman spectrum will be retained in the scale graph without loss. The 2-D data in the scale map domain is related to Raman shift and intensity. This 2-D DCNN model will be trained with several datasets and subsequently validated with other datasets for testing.

**Significance of Proposed Work**

In the proposed framework, WT and DCNN should overcome the deficiencies of classification Raman spectroscopy in complex noise environments. The main contributions of this paper can be summarised as follows:

1) The Raman spectrum noise in real environments is simulated, and the method of data preprocessing by wavelet transform is proposed. This method can simultaneously extract the characteristics of the Raman shift domain and intensity domain in Raman spectrum signals, and transform the original image data to a 224×224×3 multi-resolution scale map.

2) A DCNN is proposed as the back-end classification framework. It extracts features from multi-resolution scale maps, accelerates the training of neural networks and generates end-to-end DCNN classifiers without gradient explosion and over fitting. In this framework, the performance of ML classifiers and DCNN classifiers are evaluated according to the precision, recall rate and test accuracy.

3) The proposed framework is applied to the classification of a mixed noise Raman spectrum. The performance of the proposed framework is verified on a mixed high noise Raman spectrum data set. The accuracy, precision and robustness of the proposed framework are tested on the data set. The experimental results show that compared with NB, SVM, RF, and K-NN, the proposed framework has better classification accuracy and stronger robustness.

## II. Materials and method

In this section, a new framework based on a WT and a 2-D DCNN will be described. Fig. 1 shows a diagram of the proposed framework, which consists of three stages. The first stage is data preparation. The input of one-dimensional Raman spectrum will be pre-processed by adding noise. Noisy Raman data sets were created and grouped into training and test groups, and converted into 2D scale map data using WT. The second stage is the training of the classifier. The training data set will be applied to a ML and 2-D DCNN algorithm, and the classifier based on ML and 2-D DCNN will be obtained from this stage. Finally, the noisy Raman data sets with different noise levels are tested with different classifiers and compared with the traditional ML classifiers. At present, the primary performance of DCNN is based on the research of a two-dimensional classifier. The following content will be explained in detail from the data set generation, wavelet transform and deep neural network.

### A. MATERIAL

RRUFF is a complete collection of high-quality Raman spectroscopy databases composed of 4051 well-defined minerals [1]. Raman spectrometers obtained these Raman spectra with laser wavelengths of 532 nm and 780 nm. In this paper, we chose the Raman spectrums of Actinolite, Albite, Forsterite, Grossular, and Marialite as the noiseless spectrum datasets. In the RRUFF database, there are at most 13 original Raman spectrums of each material. In a practical environment, two noise types, which are baseline background noise (BBN) and additive white Gaussian noise (AWGN), are unavoidable. Hence we developed more Raman datasets by adding BBNs and AWGNs to these original Raman spectrums. These datasets are prepared for training and verifying the accuracy and robustness of the classifiers.

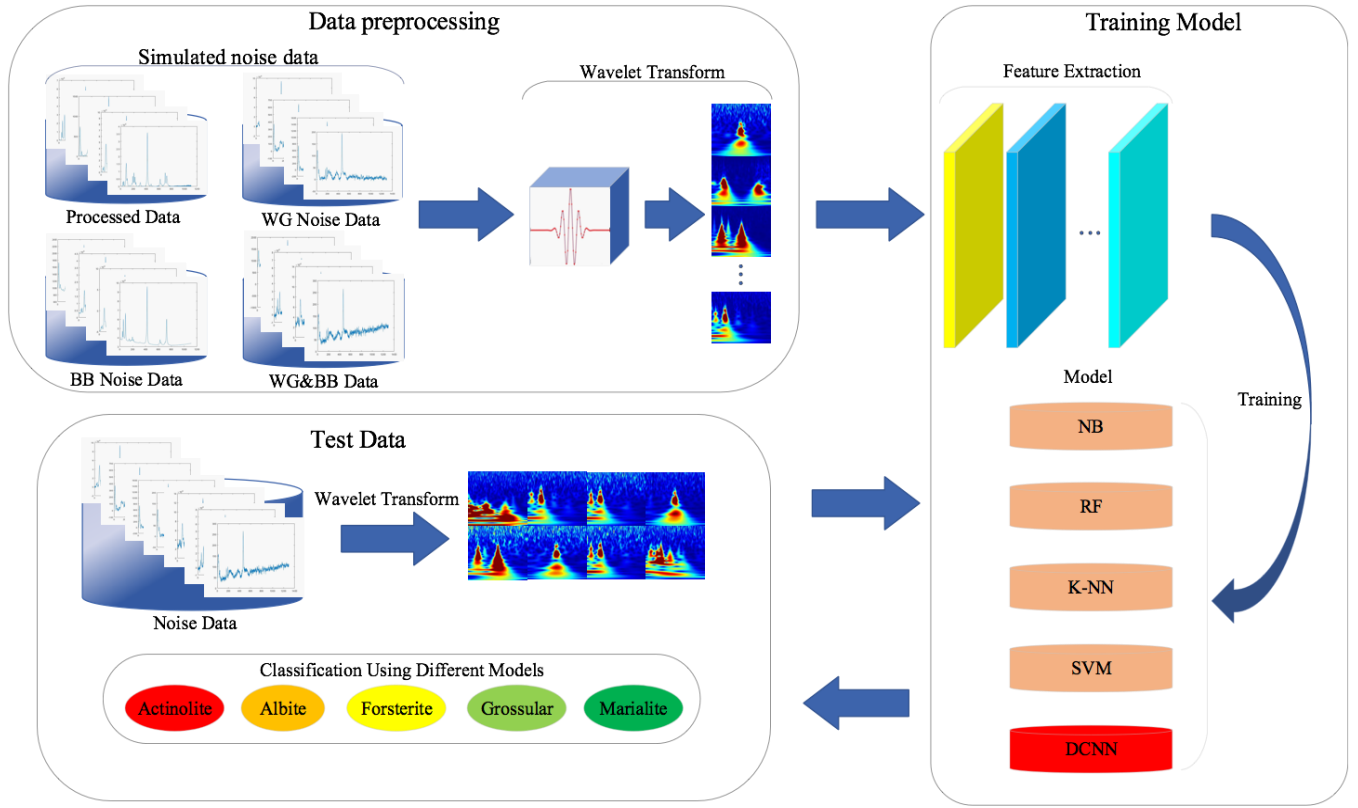

**FIGURE 1.** DNN framework based on wavelet transform.

For BBNs, several noise patterns were created. This is achieved by formulating BBN spectrums with a summation of multiple sinusoidal functions. The number of valleys in the BBN contaminated Raman spectrum depends on the number of sinusoidal functions. The BBN patterns were randomly determined by the number of sinusoidal functions, the position and amplitude of the valley peak, and the width of each valley. Different baselines are equivalent to the fluorescence noise and the other kinds of shot noise from electronic devices. For AWGN, noise signals for different noise powers were created. Both noise signals, BBN and AWGN, are added to the noiseless Raman spectrum. The Signal-to-noise ratio (SNR) is used as a parameter to quantify the noise. Theoretically, SNR is defined by:

$$SNR = 10 log_{10} \frac{P_s}{P_n} \qquad (1),$$

where $P_s$ is the signal power and $P_n$ is the noise power. In this paper, noisy Raman signal datasets with SNR of 30 dB to 80 dB are used as the training datasets while the noisy Raman signals with SNR of 1 dB to 30 dB are used as the test datasets.

Fig. 2 shows five noiseless Raman spectrums of Actinolite retrieved from the RRUFF database. As seen in Fig. 2, the spectrum footprints are quite different, even though they were obtained from a single material. This phenomenon occurs naturally for spectrum patterns of all other materials. For Raman spectroscopy application, it is necessary to train a classifier with a sufficient number of input datasets for each material. In practice, the environment for Raman spectrum sensing is noisy. This poses a challenge for developing a classifier. Fig. 3 shows 1-D Raman spectrums of Actinolite contaminated with AWGNs. Fig. 4 shows 1-D Raman spectrums of Actinolite corrupted with different BBN patterns and Fig. 5 shows 1-D Raman spectrums contaminated with both BBNs and AWGNs.

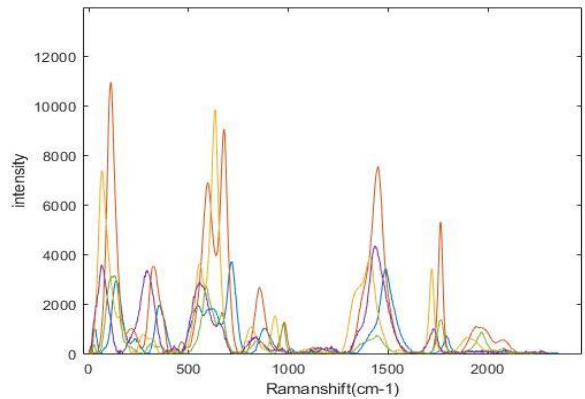

**FIGURE 2.** Original Raman spectrums of Actinolite.

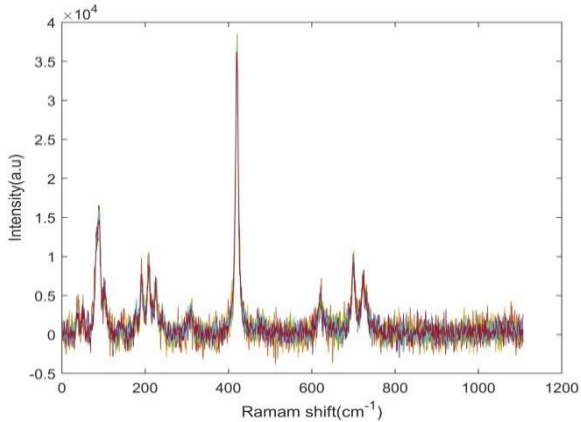

**FIGURE 3.** AWGNs contaminated Raman spectrum of Actinolite.

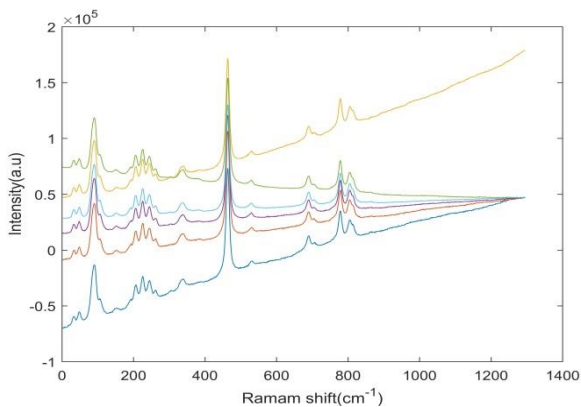

**FIGURE 4.** BBNs contaminated Raman spectrum of Actinolite.

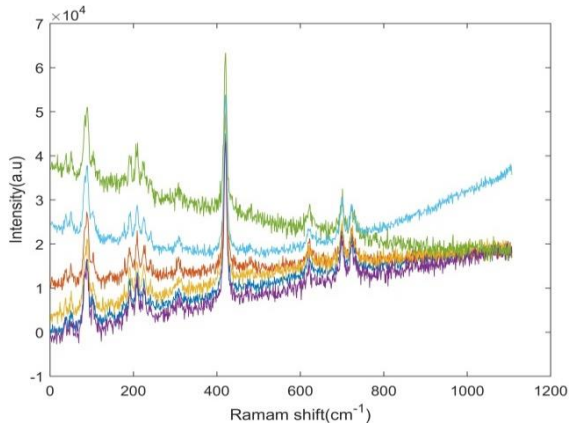

**FIGURE 5.** BBNs and AWGN contaminated Raman spectrum of Actinolite.

Table 1 shows the number of spectrum datasets applied in this paper. Five materials, Actinolite, Albite, Forsterite, Grossular, and Marialite, are listed. A total of 60 original Raman spectrums were retrieved from the RRUFF database. In this paper, we consider these original spectrums as the noiseless Raman datasets. The details of the original spectrum of each material are listed in Table 1. To train the algorithm, we created more datasets by adding BBNs and AWGNs to the noiseless spectrums. This led to 13,894 datasets of noisy spectrums. The number of datasets of each noisy spectrum class is shown in Table 1.

**TABLE 1.** Composition of Raman spectrum dataset.

| Class | Type | Number of spectrum dataset | |
| --- | --- | --- | --- |
| | | Original | Noisy |
| **1** | Actinolite | 11 | 2534 |
| **2** | Albite | 13 | 2993 |
| **3** | Forsterite | 13 | 2993 |
| **4** | Grossular | 13 | 2993 |
| **5** | Marialite | 10 | 2381 |

### B. WAVELET TRANSFORM

There are many methods to extract signal features. Fourier transform (FT) is a well-known method for extracting signal information and describing it in a spectrum domain. Since FT can effectively extract information for a stationary signal [26], [27], it is not suitable for some cases, for example, non-stationary signals, short-intervals or transient signals. Non-stationary signals such as Electroencephalography and Electrocardiography signals [28], [29] can be analysed using short-time FT [30], which divides a whole time-domain signal into several short time windows and performs FT. For time-varying non-stationary signals, high frequency is suitable for a small window and low frequency for a large window. However, the selection of the window size should be chosen carefully. Time-varying non-stationary signals processed by fast FT can only obtain the frequency components of a signal, but it does not know when the components appear. Therefore, two signals with massive time-domain difference may have the same spectrum. In other words, the FT method is not suitable to deal with the non-stationary signal, such as the Raman spectrum. For Raman spectroscopy applications, measured Raman spectrum usually fluctuates. There are spectrum shiftings and spectrum peak oscillation due to environmental noise.

A WT was proposed to deal with a non-stationary signal and a noisy environment. Unlike the basis function of FT which uses a trigonometric function, WT provides a new set of mathematic functions as a basis function. These basis functions are localised in both the time and spectrum domain. This leads to avoiding the Gibbs effect and achieve orthogonalisation. Hence better reliable and detailed time-scaled signal information is obtained than with the FT [11], [13], [31]. Theoretically, WT for a signal $x(t)$ is performed by,

$$X(a,b) = \frac{1}{\sqrt{a}} \int_{-\infty}^{\infty} \psi\left(\frac{t-b}{a}\right) x(t) dt, \quad (1),$$

where $\psi(\cdot)$ is the wavelet basis function, and $a$ and $b$ are the scale and translation variables of WT. It is shown in (1) that $a$ and $b$ control the dilation and the translation of $\psi(\cdot)$.

We propose using WT for Raman spectroscopy application. A Raman signal is transformed into a wavelet domain. In this paper, the Morlet wavelet with a centre frequency of 1 is used to transform the signal. A large number of centre frequencies are obtained by scale transformation where a series of basis functions in different intervals are obtained by Raman displacement. They are integrated with the product of a particular segment (corresponding to the interval of the basis function) of the original signal, respectively. The frequency corresponding to the extreme value is the frequency contained in this region of the original signal. WT can not only avoid the Gibbs effect but also realise orthogonalisation. We will divide the simulated Raman spectrum data into the training and testing sets. Subsequently, we will analyse them with wavelet multi-resolution [32]. After performing WT, the noise will generate a white fringe area. The higher the noise in the signal, the denser the white stripe area will be. The noise of the signal is positively correlated with the range of the stripe area.

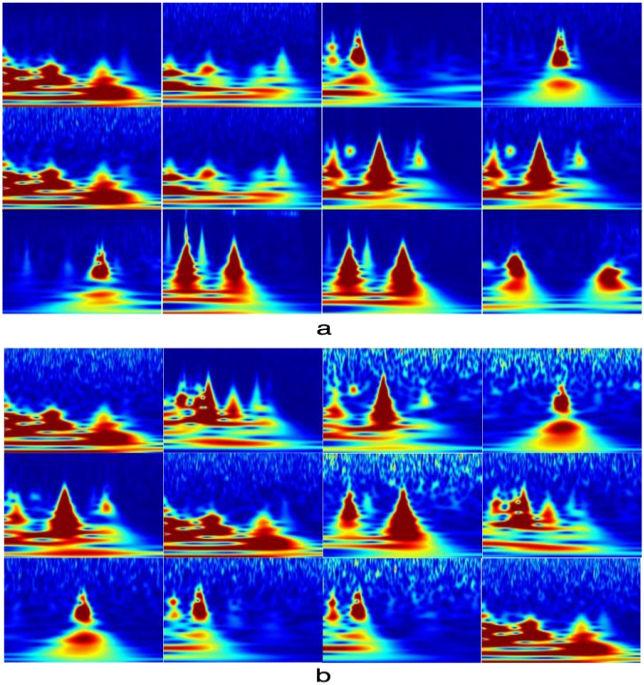

**FIGURE 6.** Scale map of Raman spectrum after WT.

## C. DEEP CONVOLUTION NEURAL NETWORK (DCNN)

Fig. 7 shows the proposed 2-D DCNN [33]. With the signal input size of 2-D DCNN of 224×224 pixels, the WT output will be 224×224 pixels. The hidden layer in the proposed

DNN consists of a set of activation functions, a full connection layer and a pooling layer. A 7 ×7 convolution kernel and 3 ×3 matrix convolution are applied [34]. The kernel convolution size is chosen such that the speed and accuracy of the feature extraction process are obtained. Due to the large-scale convolution kernel in the extraction of features, noise is inevitably introduced. Using a small convolution kernel in feature learning can not only reduce the error probability but also avoid errors caused by a large number of calculations. The convolution operations are defined by

$$Y^i = f\left(b^i + \sum_j k^{ji} * x^i\right) \quad (2),$$

where $x^i$ is the input Raman scale map, $Y^i$ is the output characteristic diagram, $*$ is the convolution symbol, $k^{ji}$ is the convolution kernel between the characteristic graphs $i$ and $j$, and $b^i$ is the $i^{th}$ weight bias. The function f($\cdot$) represents the activation function.

In this paper, 64 filters of size $7 \times 7$ are used for the analysis of the Raman spectrogram. The Relu function was chosen as the activation function of each convolution layer [34] for its handling of overfitting in the DCNN model. The function reduces the interdependence of parameters, produces a sparse neural network model, which in turn reduces the overfitting problem [35]. The Relu function introduces the nonlinear relationship to the input of the neural node $x$, which is defined by,

$$f(x) = \begin{cases} x & \text{if } x \geq 0 \\ 0 & \text{if } x < 0. \end{cases} \quad (3)$$

The pooling layer down-samples subsequently use the pooling filter to get the maximum value from the input. Each neuron pool in the output $Y_i^j$ map is on the non-overlapping region of the input $x_i$. Maximum pooling is defined by

$$Y_i^j = max\{x_{i,m}^j\}. \quad (4)$$

With the increase of the depth of NN shown in Fig 7, the parameter change of the former layer in the process of training affects the change of the latter layer (because the output of the former layer is the input of the latter). This effect increases as the network depth increases, which causes the problem of difficult network training and fitting. To solve this problem, a batch normalisation (NB) layer is added to improve the distribution of output features in the hidden layer and the convergence speed during the training state [36]. This is defined by,

$$\hat{x}^{(i)} = \frac{x^{(i)} - E[x^{(i)}]}{\sqrt{Var[x^{(i)}]}} \quad . \quad (5)$$

From (5), $\hat{x}^{(i)}$ represents the output of the BN layer and $i$ is the dimension of the feature scale map and is equal to 2. $E[x^{(i)}]$ is the mean and $Var[x^{(i)}]$ is the variance of input. The additional layer shown in Fig. 7 obtains the output features from the Relu layer and the feature map of the BN layer. This layer combines and passes to the following NN.

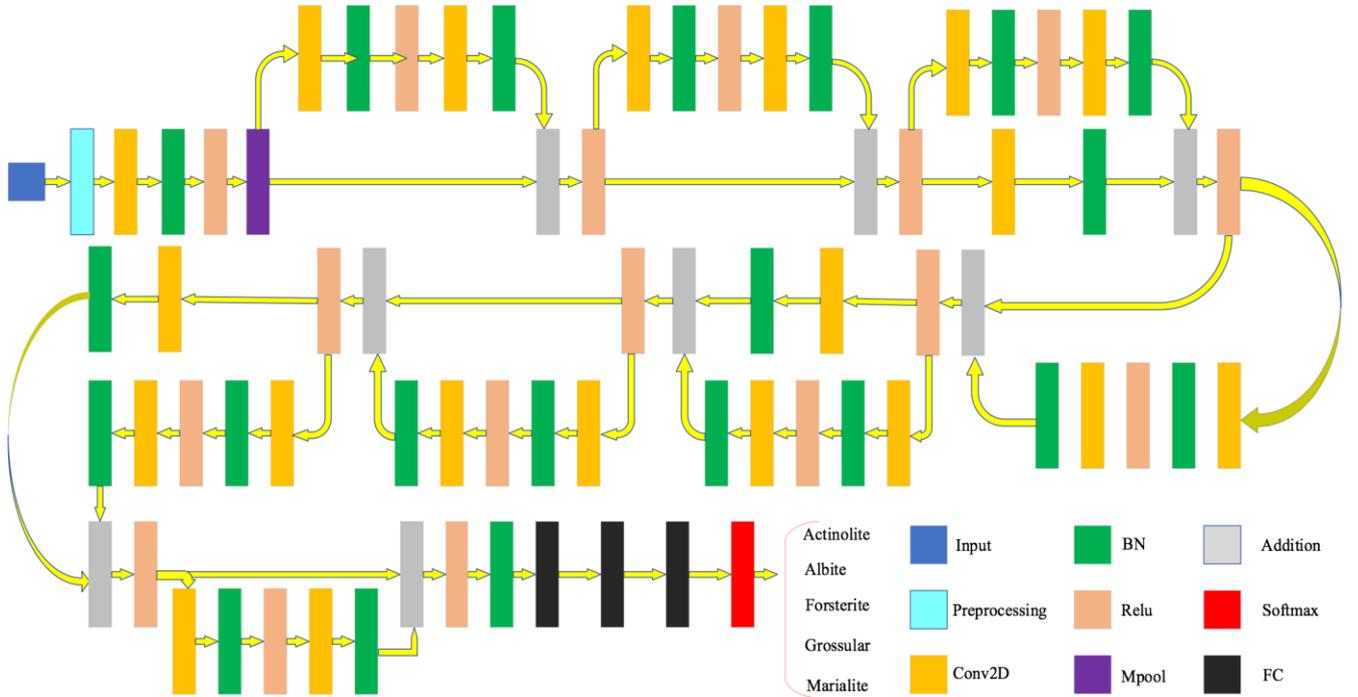

**FIGURE 7.** A simple two-dimensional deep convolution neural network model.

Shown in Fig. 7, a cascaded network of three full connection layers (FC) obtains the output from the BN layer. Each FC layer is defined by,

$$Y_i = \sum_i w_{ii'} x_i + b_{i'} \quad (6),$$

where $x_i$ represents the input of the FC layer, $w_{ii'}$ represents the weight matrix, $b_{i'}$ represents the bias and $Y_i$ represents the output of each FC layer. In short, the feature scale-map datasets are placed on one dimension, and the nonlinear problem will be solved by the multi-layer FC connection [22]. The output of the multi-layer is applied to the Softmax function to produce the classification output. This function will provide the value range between 0.0-1.0.

## III. DATA PREPARATION AND EXPERIMENT

In this work, we focus on multiple methods. Traditional ML classifiers and DCNN classifiers have great similarities in data preprocessing and feature extraction. The ML algorithm has achieved superior accuracy in the field of classified Raman spectrum [18]–[21]. In order to choose the best classifier to save time in future applications, the typical traditional ML classifiers of NB, RF, k-NN, and SVM were selected for comparison with DCNN classifiers [1]. The performances of these ML algorithms were compared with that of the DCNN network proposed in Fig. 7. The concept of cross-entropy was introduced to evaluate the loss function of the proposed DCNN network. Let $L$ represent the loss mean of output, which is defined by

$$L = -\frac{1}{N} \sum_{i=1}^{N} y^{(i)} \log \hat{y}^{(i)} + (1 - y^{(i)}) \log(1 - \hat{y}^{(i)}) \quad (7),$$

where $\hat{y}^{(i)}$ represents the feature of the Raman spectrum scale map of neuron output, $y^{(i)}$ is the corresponding target output, $N$ is the total number of training data, and the summation operator is performed on all training inputs.

## IV. RESULTS AND DISCUSSION

### A. RESULTS

The first step is to test and compare the performances among the chosen ML classifiers. In the second step, the experiment focuses on the performance analysis of the 2-D DCNN classifier.

#### 1) TRADITIONAL ML CLASSIFIERS

Fig. 8 evaluates the performance of four ML classifiers using three sets of data sets: GN, BB and GB according to the Precision, Recall, F-measure, and Test accuracy [37], [38]. The accuracy of the SVM classifier in all scenarios (BB, GN and GB) was 4.5%, 2.5% and 0.5% higher than other ML classifiers, respectively. Secondly, in BB scenarios, the accuracy of these classifiers is better than 95%. This shows that these ML classifiers are robust to BB noise. Finally, these ML classifiers are very useful for GN and GB scenarios.

#### 2) PROPOSED CLASSIFIER

The DCNN network extracts the features of the scale graph of the Raman spectrum. Due to the backpropagation error in the whole training process, the weight and deviation of the DCNN model tend to a stable range, thus improving the training accuracy. To evaluate the model, we evaluated it from the following measures:

• Training and validation accuracies.
• Precision, Recall, F-measure, and classification statistics of category forecasts [37,38].
• Testing accuracy.

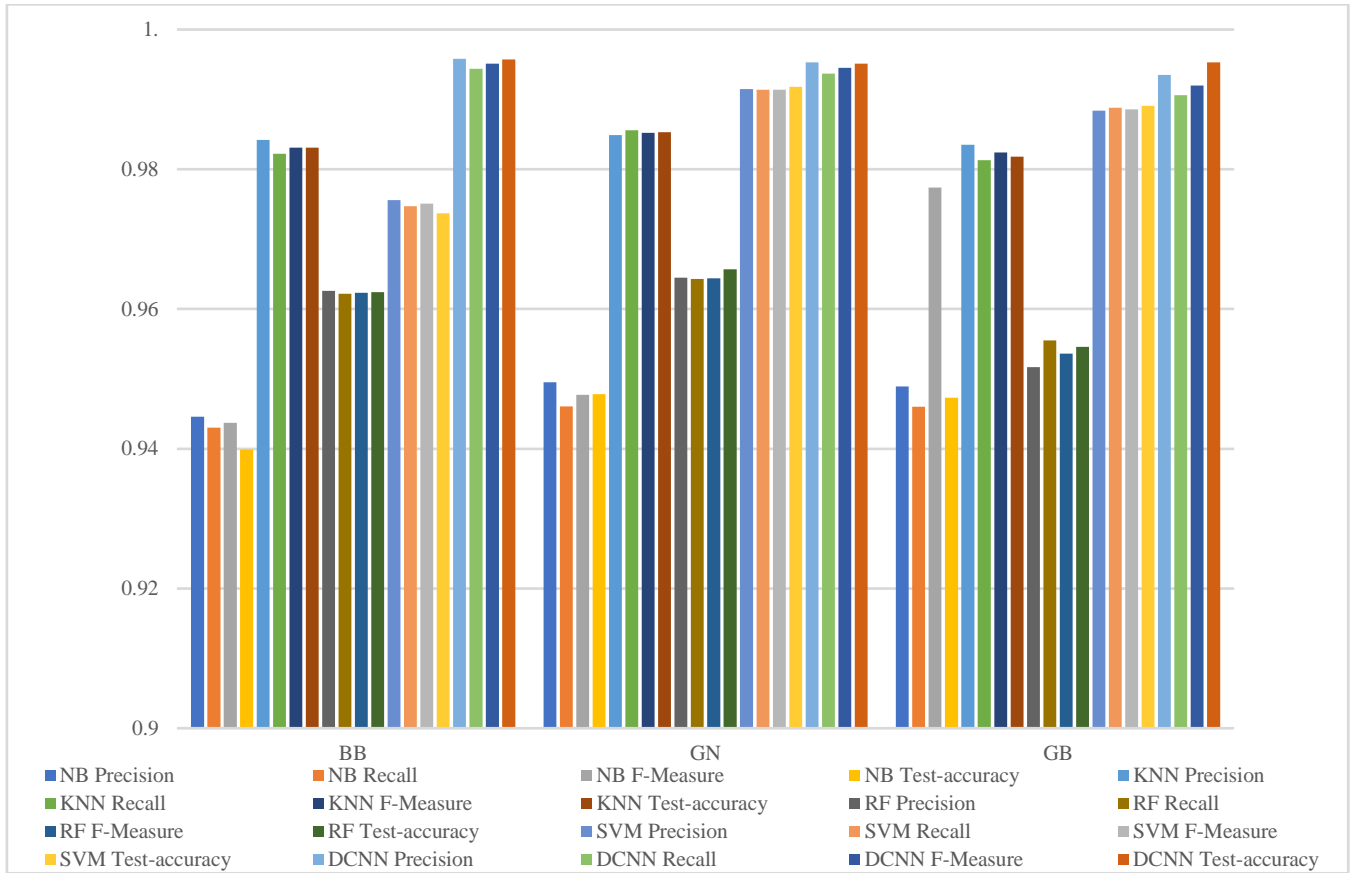

**FIGURE 8.** The proposed framework analyses the classification results of different data sets.

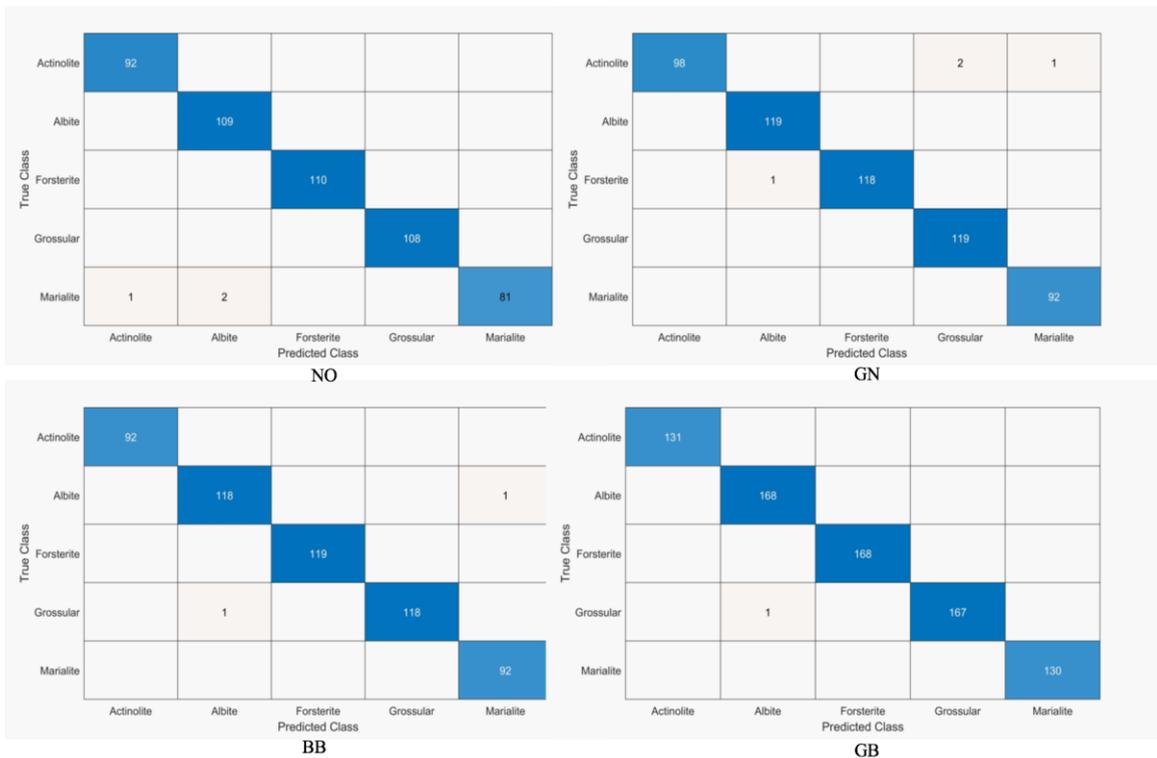

**FIGURE 9.** The confusion matrix obtained by testing different data sets, the horizontal axis represents the predicted label, and the vertical axis represents the actual label.

• Confusion matrix as a holistic measure of a classifier.

Fig. 8 details the parameters of the three evaluation indicators obtained from 2-D DCNN. The Precision, Recall and F-Measure value reached 99.4%, 99.23% and 99.3%, respectively. The training and verification accuracy is better than 99.6%, while the testing accuracy is better than 99.5%. The confusion matrix clearly shows the classifier performances and pinpoints the error occurring from each dataset group. All kinds of predicted tags are similar to real tags, with an accuracy of 99.2%, as shown in Fig .9.

### B. Discussion
According to Fig. 8, the experiment counted the use of different classifiers to test the Raman spectrum scale map under different noise scenarios. The experiment compares the ML classifiers and the DCNN classifier according to the evaluation indicators (precision, recall, F-measure, and test-accuracy). The results shown in Fig. 8 shows that the best ML classifiers under BB, GN and GB environments are K-NN, SVM and SVM, respectively. However, the DCNN classifier outperforms the ML classifiers for all noise conditions. In the BB noise scenario, the precision, recall, F-measure, and test-accuracy of the DCNN classifier were 1.1%, 1.2%, 2%, and 1.2% higher than those of K-NN. In the GN noise scenario, the precision, recall, F-measure, and test-accuracy of the DCNN were 0.4%, 0.2%, 0.3%, and 0.4% higher than those of SVM. Finally, the precision, recall, F-measure, and test-accuracy of the DCNN classifier were 0.5%, 0.2%, 0.4%, and 0.6% higher than those of SVM. To verify the accuracy of the classifier again, we tested the Raman spectrum signal (GB) with 10dB-30dB SNR from the simulated data set and tested the two classifiers. Accuracy results in Table 4 show that the proposed method is 9%, 3.5%, 1%, and 0.5% better than NB, RF, K-NN, and SVM classifiers, respectively.

**Table 4. ML classifier and DCNN classifier test for the Raman spectrum signal (GB) under 10dB-30dB SNR.**

|  | NB | RF | K-NN | SVM | DNN |
|---|---|---|---|---|---|
| Accuracy(%) | 0.8638 | 0.9217 | 0.9419 | 0.9516 | 0.9559 |

To verify the robustness of the classifier, we tested the classifiers with a noisy processed Raman scale map in the GN condition of 0dB-30dB SNR noise. The accuracy performances of the 4 ML and 2-D DCCN classifiers are shown in Fig. 10. The accuracy rate of the ML classifiers is shallow in the high noise environment, and it remains below 65% in the 0dB-14dB noise environment. On the contrary, the DCNN classifier starts to get more than 90% accuracy in a 4dB noise environment. This clearly shows that the accuracy of the DCNN classifier in the high noise environment is much better than the ML classifiers. With the increase of SNR, the accuracy of the ML classifiers start improving. However, the accuracy of a DCNN classifier already achieves above 97% at 14dB SNR.

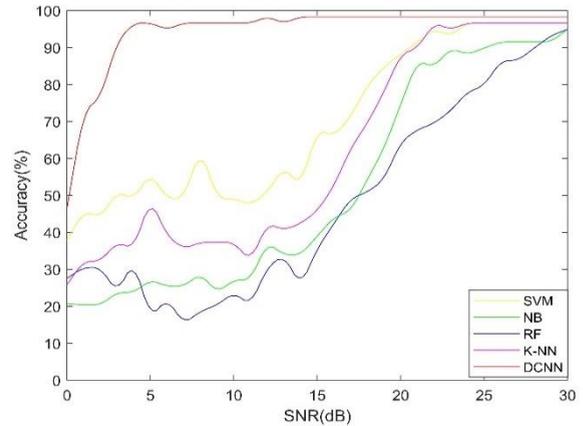

**FIGURE 10. Classification accuracy of baseline free data sets of mixed Gaussian noise by various methods.**

Fig. 11 shows the accuracy performance of both types of classifiers for the GB scenario. Among the ML classifiers, the SVM classifier model is better than other ML classifiers in this scenario. However, the DCNN classifier outperforms all ML classifiers. Even in a low SNR region, the accuracy of the DCNN classifier is better than 90%. It can be concluded that the proposed DCNN classifier is more robust than the previous ML classifiers in all noise conditions.

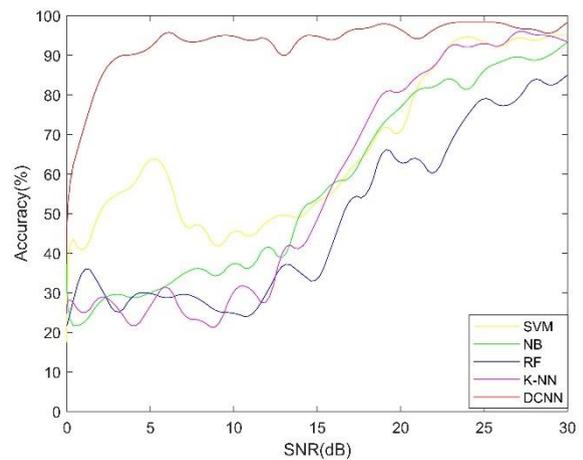

**FIGURE 11. Classification accuracy of baseline data set of mixed Gaussian noise by various methods.**

## V. Conclusion
We have proposed a new framework based on wavelet transform and deep neural network for Raman spectroscopy. The framework consists of two main engines, a wavelet transform front-end and a classifier back-end. The wavelet engine extracts the characteristics information of the Raman shift and intensity domains in the multi-resolution scale map dataset. This dataset is fed into the classifier back-end. Under three noises scenarios, which include baseline background noise, Gaussian noise, and background baseline and Gaussian noises, the DCNN classifier has been found to be the optimum classifier for the proposed framework among four machine learning classifiers: NB, RF, K-NN, and SVM. The key statistical measures of precision, recall, test-accuracy, and F-measure obtained from the deep learning network were excellent and worked better than those obtained from the machine learning algorithms. In terms of noise robustness, the framework with the DCNN classifier shows superior performance than the others. Accuracy of 90% at 3 dB SNR was achieved with the deep learning classifier while NB, SVM, RF, and K-NN require SNR of 27 dB, 22 dB, 27 dB, and 23 dB SNR, respectively. We believe that the proposed classification framework has great potential in practical Raman spectroscopy applications and also for other classification applications such as EEG, ECG and specially mixed noise signals. Our next work is to implement the framework in a low-cost portable Raman spectroscopy device.

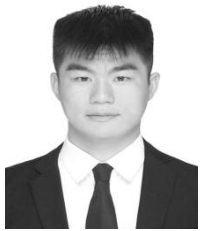

**Liangrui Pan** was born in Anhui, China in 1997. In 2019, he obtained a bachelor's degree from Anhui Polytechnic University. He is pursuing a master's degree in electrical engineering at Prince Songkhla University in Thailand in 2019. Member of IEEE, member of Chinese Society of Electrical Engineering. His research interests are machine learning, deep learning, and pattern recognition.

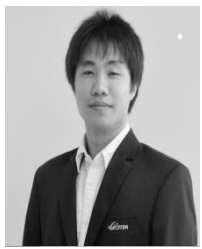

**Pronthep Pipitsunthonsan** received a bachelor's degree from Prince of Songkla University in 2010 and a master's degree in 2017. He is currently pursuing a doctorate in computer engineering. Since 2015, he has worked as a programmer at GISTDA. Its research interests are deep learning and big data.

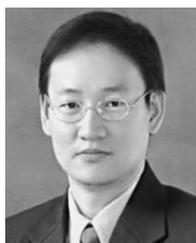

**Mitchai Chongcheawchamnan** (SM'98) was born in Bangkok, Thailand. He received the B.Eng. degree in telecommunication from the King Mongkut's Institute of Technology Ladkrabang, Bangkok, in 1992, the M.Sc. degree in communication and signal processing from Imperial College, London, U.K., in 1995, and the Ph.D. degree in electrical engineering from the University of Surrey, Guildford, U.K., in 2001. He joined the Mahanakorn University of Technology, Bangkok, as a Lecturer, in 1992. In 2008, he joined the Faculty of Engineering, Prince of Songkla University, Songkhla, Thailand, as an Associate Professor. His current research interests include microwave circuit design and microwave techniques for agricultural applications.

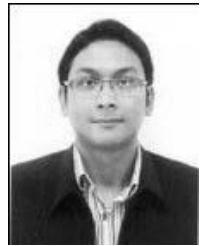

**Chalongrat Daengngam** received the B.S. in Physics from Prince of Songkla University, Thailand in 2005, M.Sc. in Nanoelectronics & Nanomechanics from University of Leeds, UK in 2006, and Ph.D. in Physics from Virginia Tech, USA in 2012. Currently, he is working as an assistant professor at Department of Physics, Faculty of Science, Prince of Songkla University. His research interests involve nonlinear optical properties of nanomaterials, photonics, and standoff Raman spectroscopy.

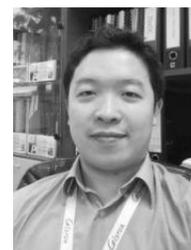

**Sittiporn Channumsin** received his undergraduate degrees in electronics engineering in 2006 at King Mongkut's Institute of Technology Ladkrabang (KMITL), Thailand, and his Master of Science in Space Technology and Planetary Exploration at University of Surrey and Ph.D. in Aerospace Engineering at University of Glasgow, United Kingdom in 2011 and 2016 respectively. He is currently a supervisor of Astrodynamics Research Laboratory (AstroLab) at Geo-Informatics and Space Technology Development Agency (GISTDA). He serves as co-editor and technical reviewer. His research and publication interests include spaceflight dynamics, space debris mitigation, machine learning and orbit and attitude control and optimisation.